\documentclass[12pt]{article}
\usepackage{amsmath,enumerate,amsfonts,color}

\usepackage{graphicx}
\usepackage{amssymb}
\usepackage{amsmath}
\usepackage{srcltx}
\usepackage{color}
\usepackage{cite}
\usepackage{url}
\usepackage{ntheorem}
\usepackage{psfrag}

\newcommand{\beadl}[1]{\begin{deqarr}\label{#1}}
\newcommand{\eeadl}[1]{\arrlabel{#1}\end{deqarr}}
\newcommand{\eead}[1]{\end{deqarr}}

\newtheorem{Theorem} {\sc  Theorem\rm} [section]

\newtheorem{Lemma} [Theorem] {\sc  Lemma\rm}

\theorembodyfont{\upshape}

\theoremsymbol{\ensuremath{\diamondsuit}}
\newcommand{\fcoco}{\small}
\theorembodyfont{\fcoco}\theoremseparator{}\theoremindent0.5cm

\theoremstyle{nonumberplain}\theorembodyfont{\fcoco}
\theoremseparator{}\theoremindent0.5cm

\DeclareFontFamily{OT1}{rsfs}{}
\DeclareFontShape{OT1}{rsfs}{m}{n}{ <-7> rsfs5 <7-10> rsfs7 <10-> rsfs10}{}
\DeclareMathAlphabet{\mycal}{OT1}{rsfs}{m}{n}

{\catcode `\@=11 \global\let\AddToReset=\@addtoreset}
\AddToReset{equation}{section}

\AddToReset{figure}{section}

\newcommand{\jlcax}[1]{}
\newcommand{\eean}{\nonumber\end{eqnarray}}

\newcommand{\newF}{\lambda}

\newcommand{\kk}[1]{}

\newcommand{\beq}{\begin{equation}}

\newcommand{\T}{\Bbb T}

\newcommand{\FS}       
                  {F}

\newcommand{\HS} 
       {H_{\mbox{\scriptsize volume}}}

{\ptc{this should be removed in the oberwolfach version}}%

\newcommand{\eeal}[1]{\label{#1}\end{eqnarray}}
\newcommand{\bed}{\begin{deqarr}}
\newcommand{\eed}{\end{deqarr}}
\newcommand{\bedl}[1]{\begin{deqarr}\label{#1}}
\newcommand{\eedl}[2]{\arrlabel{#1}\label{#2}\end{deqarr}}

\newcommand{\bel}[1]{\begin{equation}\label{#1}}
\newcommand{\bea}{\begin{eqnarray}}
\newcommand{\bean}{\begin{eqnarray}\nonumber}
\newcommand{\beal}[1]{\begin{eqnarray}\label{#1}}
\newcommand{\eea}{\end{eqnarray}}

\newcommand{\dist}{\mathrm{dist}}

\newcommand{\qed}{\hfill $\Box$ \medskip}

\newcommand{\be}{\begin{equation}}
\newcommand{\eeq}{\end{equation}}
\newcommand{\ee}{\end{equation}}
\newcommand{\beqa}{\begin{eqnarray}}
\newcommand{\eeqa}{\end{eqnarray}}
\newcommand{\beqan}{\begin{eqnarray*}}
\newcommand{\eeqan}{\end{eqnarray*}}
\newcommand{\ba}{\begin{array}}
\newcommand{\ea}{\end{array}}

\newcommand{\const}{\mbox{\rm const}} 

\newcommand{\mnote}[1]
{\protect{\stepcounter{mnotecount}}$^{\mbox{\footnotesize
$
\bullet$\themnotecount}}$ \marginpar{
\raggedright\tiny\em
$\!\!\!\!\!\!\,\bullet$\themnotecount: #1} }

\newcommand{\warn}[1]
{\protect{\stepcounter{mnotecount}}$^{\mbox{\footnotesize
$
\bullet$\themnotecount}}$ \marginpar{
\raggedright\tiny\em
$\!\!\!\!\!\!\,\bullet$\themnotecount: {\bf Warning:} #1} }

\newcommand{\R}{\mathbb R}

\newcommand{\eq}[1]{(\ref{#1})}

\newcommand{\ptc}[1]{\mnote{{\bf ptc:}#1}}

\newcommand{\Hess}{\mathrm{Hess}\,}

\newcommand{\Ric}{\mbox{\rm Ric}}

\newcommand{\beqar}{\begin{deqarr}}
\newcommand{\eeqar}{\end{deqarr}}

\newcommand{\beaa}{\begin{eqnarray*}}
\newcommand{\eeaa}{\end{eqnarray*}}

\newcommand{\tr}{\mbox{tr}}

\def\RR{{\mathbb R}}

\def\FS{{\mathfrak S}}

\newcounter{marnote}

\def\divop{{\rm div}}
\def\dist{\rm dist}

\def\stmetr{{\mathfrak{g}}}

\begin{document}

\title{Ghost points in inverse scattering constructions of  stationary Einstein metrics}
\author{Piotr T. Chru\'{s}ciel\thanks{University of Vienna}$\ $ and Luc
Nguyen\thanks{OxPDE, Mathematical Institute, Oxford}}
\maketitle

\begin{abstract}
We prove a removable singularities theorem for stationary
Einstein equations, with useful implications for constructions
of stationary solutions using soliton methods.
\end{abstract}

\section{Introduction}

The soliton technique has proved  to be a very effective tool
for constructing stationary black holes in five dimensions, see
e.g.~\cite{Pomeransky,PS,EF,EmparanReall,ERWeyl}. The method is
used to construct singular solutions of harmonic-map type
equations. One then needs to make sure that the singularity
structure of the resulting harmonic map is compatible with a
smooth geometry of the associated space-time.

Proceeding in this way, in their ingenious construction of
Black Saturns~\cite{EF}, Elvang and Figueras introduce a
singular point
$$
 \alpha_1:=(\rho=0, z=a_1)
$$
on the boundary
$\{\rho =0 \}$ of the Weyl coordinates domain $\{\rho \ge 0,
z\in \R\}$, and fine-tune certain constants to ensure that the
metric functions remain uniformly bounded near $\alpha_1$. Now,
the resulting metric ends up being a rational function of
$$
 R_1:= \sqrt{\rho^2 + (z-a_1)^2}
 \;.
$$
This leads to a potential problem because $R_1$ is \emph{not}
differentiable at $\alpha_1$, and therefore the
differentiability of the metric functions at $\alpha_1$ is not
apparent.

We will refer to such points as \emph{ghost points}, as their
occurrence does not seem to be related in any obvious way to
desirable geometric properties of the resulting space-time such
as end points of horizons or fixed points of the isometry
action, compare~\cite{Harmark,HYHighDim,HY}.

Closer inspection~\cite{CES} of the Black Saturn metric near
$\alpha_1$ shows that, with the choice of free constants that
makes the metric functions bounded,  all the metric functions
can be rewritten as rational functions of $R_1^2$; since this
last function is smooth, smoothness of the metric near
$\alpha_1$ becomes obvious. The calculations required to
establish this fact turn  out to be rather heavy, requiring
quite a bit of effort to coerce  {\sc Mathematica} to produce
the result. We emphasize that the result is non-trivial and
requires non-obvious factorisations and cancellations of
odd-order polynomials in $R_1$.

A similar trick of introducing ghost points has been used in
other related
constructions~\cite{ElvangRodriguez,EvslinKrishnan,Izumi,YazadjievBS,Yazadjiev2BR}.
The question then arises,  whether one needs to redo the
calculations of~\cite{CES} case by case, or   there is a
general mechanism which guarantees that ghost points are smooth
points for the resulting metric.

The object of this note is to show that smoothness of the
metric at such points is a consequence  of the stationary
Einstein equations with matter fields, without the need to
assume more Killing vectors. This can be roughly stated as
follows, see Theorem~\ref{T22VI0.5} for a precise version:

\begin{Theorem}
 \label{T22VI0.2}
Singularities of Lipschitz continuous stationary Einstein
metrics located on timelike submanifolds of codimension $m\ge
2$ are removable.
\end{Theorem}

We discuss in somewhat more detail in Section~\ref{s22VI10.GP}
how this theorem takes care of the ghost point problem.

\section{Stationary Einstein equations}
 \label{s22VI10.SEE}

We consider a time-independent metric in a space-time of dimension
$n+1$. Since the problem we  address is purely local, we assume that
the space-time metric functions $\stmetr_{\mu\nu}$ are   given
in local spatial coordinates ranging over a ball $B(R)\subset
\R^n$, $n\ge 2$, of radius $R$ centred at the origin, and the
prospective singularities lie along a smooth
submanifold
$$
 \Sigma\equiv\Sigma^{n - m}
$$
of $B(R)$ of
codimension $2\leq m \leq n$, with either $\partial \Sigma =
\emptyset$ or $\partial\Sigma \subset
\partial B(R)$.
We set
$$
B^*(R):= B(R)\setminus \Sigma
 \;.
$$

In adapted coordinates the metric can be written as
\begin{equation}
\stmetr = -V^2(dt+\underbrace{\theta_i
dy^i}_{=\theta})^2 + \underbrace{g_{ij}dy^i dy^j}_{=g}
	\;,\label{gme1}
\end{equation}
where $\partial_t$ is (stationary) Killing, i.e.
\begin{equation}
\partial_t V = \partial_t \theta = \partial_t g=0
	\;.\label{gme2}
\end{equation}

We allow matter fields $\varphi = (\varphi^A )$ with energy-momentum
tensor that depends upon $\stmetr$, $\partial \stmetr$, $\varphi$ and
$\partial \varphi$. For simplicity we assume that the
$\varphi^A$'s transform as scalars or tensors under coordinate
changes, and that the stationary matter field equations
constitute a tensorial system of the form
\bel{22VI0.1}
 \Delta_{g} \varphi = F(\stmetr,\partial \stmetr, \varphi, \partial \varphi) \text{ in } B^*(R)
  \;,
\ee
though a wider class of more general elliptic systems can be
easily incorporated in our analysis. We note that (linear) electromagnetic fields, for example, satisfy this assumption in Lorenz gauge.

The  Einstein equations with (possibly zero) cosmological
constant $\Lambda$ for a metric satisfying \eq{gme1}-\eq{gme2}
read (see, e.g., \cite{Coquereaux:1988ne} or \cite{YvonneBook})
\newcommand{\mydiv}{\mbox{div}}
\begin{eqnarray}\label{mainequation}
\left\{\begin{array}{l}
V\,\Delta_g V = -\frac 1{4} |\newF|_g^2 + T_{00} - \Big(\frac{n+1}{n-1}\Lambda - \frac{\tr_g(T)}{n-1}\Big)V^2
	\;,\\
\divop_g (V \newF) = 2V\Big[T_0 - \Big(\frac{n+1}{n-1}\Lambda - \frac{\tr_g(T)}{n-1}\Big)V^2\,\theta\Big]
	\;,\\
\Ric(g)- V^{-1}\Hess_gV = \frac{1}{2V^{2}}\newF\circ \newF + T_g
 + \Big(\frac{n+1}{n-1}\Lambda - \frac{\tr_g(T_g)}{n-1}\Big)g
	\;,
\end{array}\right.
\end{eqnarray}
where $T_g:=T_{ij}dx^i dx^j$, $T_0 = T_{0i}\,dx^i$, and
$$\newF_{ij}=-V^2(\partial_i \theta_j - \partial_j
\theta_i)\;,\;\;\;(\newF\circ \newF)_{ij}=\newF_i{^k}\newF_{kj}\;.
$$
Altogether the Einstein-matter field equations, which are
supposed to hold in $B^*(R)$, can therefore be written as
\begin{eqnarray}\label{EFE}
\left\{\begin{array}{l}
\Delta_g \varphi = F(\stmetr,\partial \stmetr, \varphi, \partial \varphi)
	\;,\\
\Delta_g V = F_1(\stmetr, \partial\stmetr ) + \tilde T_V
	\;,\\
\divop_g (d\theta) - d(\divop_g(\theta)) = F_2(\stmetr, \partial\stmetr ) + \tilde T_\theta
	\;,\\
\Ric(g) - V^{-1}\Hess_gV= F_3(\stmetr, \partial\stmetr) + \tilde T_g
	\;,
\end{array}\right.
\end{eqnarray}
for some (explicitly computable) $F_1$, $F_2$ and $F_3$ which
are polynomials in $\stmetr$, $\stmetr^{-1}$, $\partial\stmetr$
and quadratic in $\partial\stmetr$ and some $\tilde T_V$, $\tilde T_\theta$ and $\tilde T_g$ which arise from $T_{\mu\nu}$.

\medskip

We have:

\begin{Theorem}
 \label{T22VI0.5}
Under the conditions above, suppose that $(\stmetr,\varphi) \in
C^{0,\alpha}(B(R)) \cap C^2(B^*(R))$  and
$V > 0$ in $B(R)$.
Assume further that
\begin{enumerate}
 \item  either $\alpha=1$,
\item or $\frac{n-m}{n} <\alpha<  1$ and  there exists a constant $C$,
    possibly depending upon $(\stmetr,\varphi)$, such that
\begin{equation}\label{Hyp1}
 |\partial \stmetr| + |\partial \varphi| \le C\,{\dist}_{\RR^n}(\cdot,\Sigma)^{\alpha-1}
 \;,
\end{equation}
and
$$
 |T_{\mu\nu}| + |F| \le C(1 + |\partial \stmetr|^2 + |\partial \varphi|^2)
 \;.
$$
\end{enumerate}
Then
$$
 \stmetr\,, \, \varphi \in C^{\omega}(B(R/2)) \;.
$$
\end{Theorem}

The proof will use the following simple lemma, whose proof is
deferred until after the proof of Theorem \ref{T22VI0.5}.

\begin{Lemma}\label{RSEEqn}
Let $\Omega$ be an open subset of $\RR^n$ and $\Sigma \subset \Omega$ be a smooth submanifold of codimension $m \geq 2$ which either has no boundary or has boundary contained in $\partial \Omega$. Assume that $u \in W^{1,2}(\Omega)$ satisfies
\[
\partial_i (a^{ij} \partial_j u) = \partial_i g^i + f \text{ in } \Omega \setminus \Sigma
\]
in the sense of distributions for some $a^{ij} \in
L^\infty(\Omega)$, $f \in L^1(\Omega)$ and $g^i \in
L^2(\Omega)$. Then $u$ satisfies the above equation in $\Omega$
in the sense of distributions.
\end{Lemma}

\noindent{\sc Proof of Theorem~\ref{T22VI0.5}:}
By hypothesis there exist coordinates $y^i$  in which the
metric coefficients and the fields satisfy
\bel{25VI0.0}
V, \theta_i, g_{ij},\varphi \in C^{0,\alpha}(B(R))\cap C^2(B^*(R))
 \;.
\ee
Standard arguments (compare~\cite{MzH}) show that the metric is
in fact smooth away from $\Sigma$:
\bel{25VI0.1}
V, \theta_i, g_{ij},\varphi \in C^{0,\alpha}(B(R))\cap C^\infty(B^*(R))
 \;.
\ee

Since the problem is local, it suffices to establish the
desired regularity in a small ball $B(\epsilon)$ centered at a
point on $\Sigma \cap B(R/2)$, which is assumed to be the
origin. By a linear change of coordinates we can without loss of
generality assume that $g_{ij}(0)=\delta_i^j$.
Furthermore, we can also assume
that
\begin{equation}
\Sigma \cap B(10\epsilon) = \{y = (y',0): y' \in \pi(\Sigma)\}
	\label{30VI10.Luc02}
\end{equation}
for some open set $\pi(\Sigma) \subset B(10\epsilon) \cap
\RR^{n-m}$.

Most of our arguments rely on elliptic estimates. Of the four
equations in \eqref{EFE}, the last two are not manifestly
elliptic. As is well known, this issue can be cured by passing
to harmonic space-coordinates and using an appropriately
chosen time function.

Suppose, first, that $\frac{n-m}{n}<\alpha<1$.

For $\epsilon>0$  let $x^i$ be solutions of the problem
$$
 \Delta_{g} x^i = 0
  \;, \quad x^i |_{S(\epsilon)} = y^i
  \;.
$$
where $S(\epsilon):= \partial B(\epsilon)$ is a
$y^i$--coordinate sphere of radius $\epsilon$.

By~\cite[Theorem~8.34]{GT},
and elliptic regularity away from the origin,
we have $x^i \in C^{1,\alpha}(\bar B(\epsilon))\cap
C^{\infty}(B( \epsilon)\setminus\Sigma)$.  If we write
$$
 x^i  = y^i + f^i
 \;,
$$
then $ f^i$ solves the divergence-type equation
$$
 \partial_i(\sqrt{\det g} g^{ij}\partial_j f^\ell)  = -
 \partial_i(\sqrt{\det g} g^{i \ell})= -
 \partial_i(\sqrt{\det g} g^{i \ell}-\delta_i^\ell)
 \;.
$$
Let
$$f^\ell_\epsilon (x):= f^\ell(\epsilon x)
 \;,
 \quad
\psi^{i\ell}_\epsilon(x)
  := \sqrt{\det g(\epsilon x)} g^{i \ell}(\epsilon x)
	\;,
\quad
  \bar\psi^{i\ell}_\epsilon(x)
  := \psi^{i\ell}_\epsilon(x) - \psi^{i\ell}_\epsilon(0)
 \;.
$$
Then
\[
 \partial_i(\psi^{ij}_\epsilon\,\partial_j f^\ell_\epsilon)
= -\epsilon\,\partial_i\,\bar \psi^{i\ell}_\epsilon\ \text{ in } B(\epsilon)
	\;.
\]
Applying \cite[Theorem~8.33]{GT} to the equation satisfied by
$f^\ell_\epsilon$ on a ball of radius two we obtain (note that
the first term $|u|_0$ there can be discarded by the usual
argument that exploits injectivity of the Laplace equation)
\bel{22VII0.1}
 \|f^\ell_\epsilon\|_{C^{1,\alpha}(B(1))} \le C \epsilon\,\| \bar\psi^{i\ell}_\epsilon\|_{C^{0,\alpha}(B(1))}
 \le C  \epsilon^{1+\alpha}
  \;.
\ee
It thus follows that
\bel{22VII0.2}
 \frac{\partial x^i}{\partial y^j } = \delta^i _j + O(\epsilon^\alpha) \text{ in } B(\epsilon)
 \;.
\ee
The implicit function theorem shows the $x^i $'s can be used as
a coordinate system near $y^i =0$ for all $\epsilon$ small
enough. In what follows we choose some such value of
$\epsilon$.

As for the choice of a spacelike slice, we will now show that we can assume without loss of generality that
\begin{equation}\label{GaugeCond}
\divop_g (\theta) = 0 \text{ in } B(\epsilon) \text{ for sufficiently small } \epsilon
	\;.
\end{equation}
To use the freedom of defining $t$ (equivalently, to fix the
gauge freedom of $\theta$),
 we make a coordinate change of the
form $\tilde t = t + h(y)$. The metric $\stmetr$ then takes the
form
\[
\stmetr = - V(d\tilde t + \underbrace{(\theta_i - h_{,i})\,dy^i}_{=:\tilde\theta_i\,dy^i\equiv \tilde\theta})^2 + g_{ij}\,dy^i\,dy^j
	\;.
\]
To obtain \eqref{GaugeCond}, we pick $h \in W^{1,2}(B(\epsilon))$ to be a solution to
\[
\Delta_g h = \divop_g(\theta) \text{ in } B(\epsilon)
	\;, \text{ and }  h|_{S(\epsilon)} = 0
	\;.
\]
By~\cite[Theorems~8.12 and 8.34]{GT}, we have
$$h \in
 C^{1,\alpha}(\bar B(\epsilon)) \cap W^{2,2}(B(\epsilon)) \cap
 C^{\infty}(B( \epsilon)\setminus\Sigma)
 \;.
$$

Recalling \eqref{Hyp1} and rewriting the above equation for $h$
as
\[
g^{k\ell}\,\partial_k\,\partial_{\ell} h = - \partial_{\ell} h\,\partial_k (\sqrt{\det g}\,g^{k\ell}) + \frac{1}{\sqrt{\det g}}\partial_k(\sqrt{\det g}\,g^{k\ell}\theta_{\ell})
	\;,
\]
we can apply \cite[Theorem~9.11 and Lemma~9.16]{GT} to get,
\[
\partial^2 h \in L^q(B(\epsilon)) \text{ for any } 1 < q < \frac{m}{1 - \alpha}
	\;;
\]
here and in what follows the norm might depend upon $\epsilon$,
but this is irrelevant since a small epsilon has been now
fixed.  We have thus achieved \eqref{GaugeCond} with a penalty
that the derivatives $\partial \theta$ no longer satisfy a
pointwise estimate given by \eqref{Hyp1} but the weaker
estimate
\begin{equation}
\partial \theta \in L^q(B(\epsilon)) \text{ for any } 1 < q < \frac{m}{1 - \alpha}
	\;.\label{Hyp1-Alt}
\end{equation}
The above suffices for our purposes. We emphasize that $V$ and
$g$ remain unchanged under the redefinition of $\theta$,
equivalently of time, as above.

To prepare for our passing to the coordinates $x^i$, we need
some bound for the Hessian of $f^i$. For $y \in \RR^n$, we will
write $y = (y', y'')$ where $y' \in \RR^{n-m}$ and $y'' \in
\RR^m$. In view of \eqref{30VI10.Luc02}
we have
\begin{equation}
d(y) := {{\dist}}_{\RR^n}(y, \Sigma) = |y''| \text{ for } y \in B(5\epsilon) \setminus \Sigma
	\;.\label{30VI10.Luc03}
\end{equation}
Define
\begin{align*}
Q'_{s,S} &= \{y: s/4 \leq |y''| \leq 5s/4, |y| \leq S\}
	\;,\\
Q_{s,S} &= \{y: s/2 \leq |y''| \leq s, |y| \leq S\}
	\;.
\end{align*}
Then, by \eqref{Hyp1}, $f^\ell_s$ satisfies
\[
|\psi^{ij}_s\,\partial_i\partial_j f^\ell_s| \leq C\,s^{\alpha + 1}
	\text{ in } Q'_{1,s^{-1}\,\epsilon}
	\;.
\]
Thus, by \cite[Theorem~9.11]{GT} and \eqref{22VII0.1},
\[
\|\partial^2 f^\ell_s\|_{L^q(Q_{1,s^{-1}\,\epsilon/2})} \leq C\,s^{\alpha + 1}\,\big[\mathcal{H}^{n-m}(\Sigma_s)\big]^{1/q} \text{ for any } 1 < q < \infty
	\;,
\]
where $\mathcal{H}^{n-m}$ denotes the $(n-m)$-dimensional
Hausdorff measure: More precisely, we first apply
\cite[Theorem~9.11]{GT} to cubes of unit size and $f^\ell_s -
L(f^\ell_s)$ with $L(f^\ell_s)$ being the linearization of
$f^\ell_s$ at the center of those cubes, and then sum the
acquired estimates over a collection of non-overlapping cubes
covering the desired region. Because of the simple geometry of
$\Sigma_s$ (compare \eqref{30VI10.Luc02}),  the number of cubes
in each such collection is proportional to $s^{m-n}$, which is
itself proportional to the Hausdorff dimension above. Scaling
back, it follows that
\[
\|\partial^2 f^\ell\|_{L^q(Q_{s,\epsilon/2})} \leq C\,s^{\alpha - 1 + m/q} \text{ for any } 1 < q < \infty
	\;.
\]
Now if we pick $q$ such that $\alpha - 1 + m/q > 0$, we can sum
the above over dyadic rings in the tranverse direction to get
\begin{equation}
\|\partial^2 f^\ell\|_{L^q(B(\epsilon/2))} \leq C
 \text{ for any } 1 < q < \frac{m}{1-\alpha}
	\;.\label{HessfBnd}
\end{equation}

We pass now to the coordinates  $x^i = y^i + f^i$, and still
use the symbol $\stmetr$, $g$, $\theta$ and $\varphi$ for the space-time metric, the spatial metric, the shift one-form  and the matter
fields in the new coordinates.  Shifting the $x^i$'s by a
constant vector if necessary, we can assume that $x^i (0) = 0$.
Furthermore,
one has in the new coordinates
\begin{align*}
&V, \theta_i, g_{ij},\varphi \in C^{0,\alpha}(B(\epsilon))\cap C^\infty(B^*(\epsilon))
	\;.
\end{align*}

Estimate \eqref{HessfBnd} shows that
$x^i\in W^{2,q}(B(\epsilon/2))$, and by \eqref{Hyp1}, \eqref{Hyp1-Alt}, the chain rule and the
transformation law for tensors one deduces that
\begin{equation}
|\partial_x \stmetr| + |\partial_x \varphi| \in L^q(B(\epsilon/2)) \text{ for any } 1 < q < \frac{m}{1-\alpha}
	\;.\label{LqEst}
\end{equation}

In the coordinates $x^i $ the Einstein-field equations
\eqref{EFE} can be rewritten in the following form
\begin{eqnarray}\label{EFE'}
\left\{\begin{array}{l}
\Delta_g \varphi^A = F(\stmetr, \partial\stmetr, \varphi, \partial \varphi)
	\;,\\
\Delta_g V = F_1(\stmetr, \partial\stmetr) + \tilde T_V
	\;,\\
\Delta_g \theta_i = F_{(i)}(\stmetr, \partial\stmetr) - (\tilde T_\theta)_i
	\;,\\
\Delta_g g_{ij} -  \partial_i (\partial_j \log V) = F_{(i)(j)}(\stmetr, \partial\stmetr) + (\tilde T_g)_{ij}
	\;,
\end{array}\right.
\end{eqnarray}
where we have used \eqref{GaugeCond}. Using \eqref{LqEst} together with the given growth rate of $F$ and $T$, one sees that $(V, \theta, g, \varphi) \in W^{1,2}(B(\epsilon/2))$ while the right side of \eqref{EFE'} belongs to $L^p(B(\epsilon/2))$ for any $p < \frac{m}{2(1-\alpha)}$. Also, by Lemma \ref{RSEEqn}, \eqref{EFE'} is satisfied across $\Sigma$ in the sense of distribution. It is useful to write the last equation in \eqref{EFE'} as
\begin{equation}\label{EFE'last}
\Delta_g g_{ij} =  \partial_i (\partial_j \log V) + \tilde F_3(\stmetr, \partial\stmetr) + (\tilde T_g)_{ij}
	\;.
\end{equation}

To proceed, we distinguish two cases according to whether
$\alpha > 1 - \frac{m}{2n}$ or $\alpha \leq 1 - \frac{m}{2n}$.
In the former case, we apply \cite[Theorem
5.5.3(b)]{Morrey} to the first three equations of \eqref{EFE'} to assert that
$(\varphi, V, \theta) \in C^{1,\sigma}(B(\epsilon/3))$ for some
$\sigma > 0$. In particular, $\partial \log V \in
C^{0,\sigma}(B(\epsilon/3))$. Applying \cite[Theorem
5.5.3(b)]{Morrey} again to \eqref{EFE'last}, we get $g \in
C^{1,\sigma}(B(\epsilon/4))$.

In the latter case, we use \cite[Theorem 5.5.3(a)]{Morrey}. Applying this result to the first three equations in \eqref{EFE'} and then to \eqref{EFE'last} as in the previous paragraph we get $(\varphi, V, \theta, g) \in W^{1,q}(B(\epsilon/4))$ for any $1 < q < \frac{m}{2(1-\alpha) - \frac{m}{n}}$. In other words, in $B(\epsilon/4)$, \eqref{LqEst} is improved with $\alpha$ replaced by $\alpha + \big(\alpha - \frac{n - m}{n}\big)$. Repeating this process for a finite number of time, we arrive at a situation when the argument in the previous paragraph applies.

In any event, one obtains $(g,\varphi)\in C^{1,\sigma}(B(\epsilon/100))$. A standard bootstrap argument
based on Schauder estimates proves smoothness; analyticity
readily follows.

When $\alpha=1$ we replace $\alpha$ by any number in $(1 - \frac{m}{2n},1)$
and arrive to \eq{LqEst} as before. Since $T_{\mu\nu} $ is a
tensor, it gives a bounded contribution to \eq{EFE'},
leading to a metric with improved regularity as before.
Similarly $F$ is a tensor  giving a bounded contribution to
\eq{22VI0.1}, and we obtain $ (g,\varphi)\in
 C^{1,\sigma}(B(\epsilon/2))$ by the same method as above. The result follows.
\qed

To finish this section, we provide the

\medskip
\noindent{\sc Proof of Lemma \ref{RSEEqn}:} Let $\xi \in C^\infty_c(\Omega)$, we need to show that
\begin{equation}
\int_{\Omega} a^{ij}\,\partial_i u\,\partial_j \xi\,dx = \int_{\Omega} (g^i\,\partial_i \xi + f\,\xi)\,dx
	\;.\label{RSEE.01}
\end{equation}

Let $\eta$ be a smooth cut-off function on $\RR$ such that $\eta(t) = 0$ for $t \leq 1$ and $\eta(t) = 1$ for $t \geq 2$. For $\delta$ sufficiently small, define
\[
\varrho_\delta(x) = \left\{\begin{array}{l}
\eta\Big(\frac{d(x,\Sigma)}{\delta}\Big) \text{ for } m \geq 3
	\;,\\
\eta\Big(\frac{\log(-\log d(x,\Sigma))}{\log(-\log \delta)}\Big) \text{ for } m = 2
	;.
\end{array}\right.
\]
By hypothesis we have
\[
\int_{\Omega} a^{ij}\,\partial_i u\,\partial_j (\xi\,\varrho_\delta)\,dx = \int_{\Omega} (g^i\,\partial_i (\xi\,\varrho_\delta) + f\,\xi\,\varrho_\delta)\,dx
	\;.
\]
\eqref{RSEE.01} can be reached by passing $\delta \rightarrow 0$ using Lebesgue's dominated convergence theorem, Cauchy-Schwarz's inequality and the explicit form of $\varrho_\delta$. Note that when $m = 2$, we need to use
\[
\int_0^1 \frac{1}{t\,(\log t)^2}\,dt < \infty
	\;.
\]
We omit the details.
\qed

\section{Ghost points}
 \label{s22VI10.GP}

We show how Theorem~\ref{T22VI0.5} applies to stationary
solutions obtained by introducing ghost points in the solitonic
solution-generating technique. Here the space-time metric that
one wishes to construct is invariant under an abelian isometry
group $\R\times \T^{n-2}$, where the $\R$ factor represents
$t$--translations. The metric depends only upon two coordinates
$(\rho,z)$, which can be thought of as cylindrical coordinates
on $\R^3$. In this construction, if the ghost point is placed
at $(\rho=0,z=0)$, one obtains a solution of the vacuum
Einstein equations $ \stmetr_{\mu\nu}$ defined on
$$
 \{t\in \R\;,(x,y,z)\in B(\delta)\subset  \R^3\}\times \T^{n-3}
 \;,
$$
where
$$
 x= \rho \cos \varphi\;, \quad  y = \rho \sin \varphi\;.
$$
Note that one $S^1$ factor from $\T^{n-2}$ has been interpreted
as a rotation around the $z$--axis of $\R^3$. Furthermore,
there exists a neighborhood of the origin in which the metric
functions are analytic functions of $(\rho^2, z,d)$, where $d$
is the Euclidean distance to the origin in $ \R^3$:
$$
 d:= \sqrt{\rho^2 + z^2}
 \;.
$$
So the singular set
$$
\Sigma= \{x=y=z=0\}\times \T^{n-3}
$$
has dimension $ n-3$ within each slice $t=\const$, hence
codimension three.  To apply Theorem~\ref{T22VI0.5} we need to
verify that, reducing $\delta$ if necessary,
\bel{25vi0.6}
 \mbox{there exists   $\varepsilon>0$ such that $\det {}\stmetr ^{i j} >\varepsilon$  and  $
{}\stmetr _{tt}<-\varepsilon$.} \ee
Note that the function $d$ is Lipschitz-continuous, but not
differentiable. This implies that the metric functions are in
$C^\infty(B^*(R))\cap C^{0,1}(B(R))$. Theorem~\ref{T22VI0.5}
with $\alpha=1$ shows then that the metric functions are
real-analytic in a whole neighborhood of the origin of $\R^n$,
as desired.

In the case of the Black Saturn metric we have $n=4$, and the
space-dimension of the singular set is one. To verify that this
metric is analytic near its ghost point $\alpha_1=(\rho=0,
z=a_1)$, one needs to verify \eq{25vi0.6}. A direct
verification in the coordinate system used in~\cite{EF} fails,
because at this point $\stmetr_{tt}$ becomes null for all Black
Saturn metrics (indeed, $\alpha_1$ always lies on the
ergosurface for those metrics). This can be bypassed by
checking that the limit of the metric at $\alpha_1$ is
Lorentzian, and that the determinant of the matrix of scalar
products of all Killing vectors there has a strictly negative
value. This guarantees that some linear combination of Killing
vectors is timelike at $\alpha_1$, and our theorem applies.

As another application, our analysis reduces the question of
regularity of the metrics
of~\cite{ElvangRodriguez,EvslinKrishnan,Izumi,YazadjievBS,Yazadjiev2BR}
near their ghost points to showing that the components of the
metric tensor in coordinates $(x,y,z)$  have a finite limit at
the ghost points, and verifying \eq{25vi0.6} there (after
perhaps replacing $\partial_t$ by a different, timelike Killing
vector if necessary). We note that this might require tedious
symbolic algebra calculations, and our experience with the
Black Saturn metrics suggests that the checking of the timelike
character of the orbit of the isometry group  through the ghost
point might be non-trivial. In any case we have not attempted
to carry this out.

\bigskip
\noindent{\textsc{Acknowledgements:}  We are grateful to
Henriette Elvang and Pau Figueras for useful correspondence and
bibliographical advice. Hospitality and financial support of
BIRS  is acknowledged. PTC is supported in part by the Polish
Ministry of Science and Higher Education grant Nr N N201
372736. LN is supported in part by the EPSRC
Science and Innovation award to the Oxford Centre for Nonlinear
PDE (EP/E035027/1).}

\def\polhk#1{\setbox0=\hbox{#1}{\ooalign{\hidewidth
  \lower1.5ex\hbox{`}\hidewidth\crcr\unhbox0}}}
  \def\polhk#1{\setbox0=\hbox{#1}{\ooalign{\hidewidth
  \lower1.5ex\hbox{`}\hidewidth\crcr\unhbox0}}} \def\cprime{$'$}
  \def\cprime{$'$} \def\cprime{$'$} \def\cprime{$'$}
\providecommand{\bysame}{\leavevmode\hbox to3em{\hrulefill}\thinspace}
\providecommand{\MR}{\relax\ifhmode\unskip\space\fi MR }
\providecommand{\MRhref}[2]{%
  \href{http://www.ams.org/mathscinet-getitem?mr=#1}{#2}
}
\providecommand{\href}[2]{#2}

\end{document}